\documentclass[12pt]{article}
\usepackage{graphicx}
\usepackage{amsmath}
\usepackage{amsfonts}
\usepackage{amssymb}

\begin{document}

\title{Isotropic cosmologies in Weyl geometry}
\author{John Miritzis\\Department of Marine Sciences, University of the Aegean\\University Hill, Mytilene 81100, Greece}
\date{\today}
\maketitle
\begin{abstract}
We study homogeneous and isotropic cosmologies in a Weyl spacetime. We show
that for homogeneous and isotropic spacetimes, the field equations can be
reduced to the Einstein equations with a two-fluid source. We write the
equations as a two-dimensional dynamical system and analyze the qualitative,
asymptotic behavior of the models. We examine the possibility that in certain
theories the Weyl 1-form may give rise to a late  accelerated expansion of the
Universe and conclude that such behaviour is not met as a generic feature of
the simplest cosmologies.
\end{abstract}

\section{Introduction}

Recent observations regarding the evolution of the Universe have led several
authors to propose a variety of departures from the standard cosmology. These
proposals can be roughly grouped into two categories. First, there exists a
dark energy of unknown nature which is responsible for the accelerating
expansion of the Universe, (cf. \cite{pera,sast} for comprehensive reviews and
references). Alternatively general relativity requires a modification at
cosmological distance scales \cite{cdtt,chib} (for a pedagogical review see
\cite{caro}). Less explored is the idea that the geometry of spacetime is not
the so far assumed Lorentz geometry (see for example \cite{cct}). Due to its
simplicity, Weyl geometry is considered as the most natural candidate for
extending the Riemannian structure. In this geometry the covariant derivative
of the metric tensor is not zero. For example, in the spirit of the idea of
creation of the Universe from quantum fluctuations of ``nothing'', Novello
\emph{et al} \cite{nove} considered perturbations of the geometry of the form
\[
\delta\left(  \nabla_{\rho}g_{\mu\nu}\right)  =\left(  \delta Q_{\rho}\right)
g_{\mu\nu},
\]
where $Q_{\rho}$ is the gradient of a scalar field. Other motivations for
reconsidering Weyl geometry include the hope to remove the cosmological big
bang singularity.

Weyl geometry can be incorporated a priori in a theory or a posteriori, for
example as a consequence of the variational principle involved. This is the
case of the application of the Palatini method to gravitational Lagrangians of
the form $L=f\left(  R\right)  $ \cite{voli,flan}, where it turns out that
$Q_{\rho}=\partial_{\rho}\ln f^{\prime}\left(  R\right)  $. However, the
Palatini device suffers from serious problems and leads to inconsistencies
when applied to general Lagrangians for the construction of a gravity theory
(see for example \cite{buch,quer} for a thorough critique of the Palatini
device). In \cite{cmq} it was shown that a consistent way to incorporate an
arbitrary connection into the dynamics of a gravity theory is the so-called
constrained variational principle.

In this paper we study homogeneous and isotropic cosmologies in a Weyl
framework. More precisely, we explore the field equations derived with the
constrained variational principle from the Lagrangian $L=R+L_{matter},$ under
the condition that the geometry be Weylian \cite{cmq}. If the matter
Lagrangian is chosen so that ordinary matter is described by a perfect fluid
with equation of state $p_{2}=\left(  \gamma_{2}-1\right)  \rho_{2},$ we show
that for homogeneous and isotropic spacetimes, the field equations reduce to
the Einstein equations with a two-fluid source. The first fluid with equation
of state $p_{1}=\rho_{1}$ (i.e. stiff matter), stems from the Weyl vector
field and the second fluid describes ordinary matter as mentioned above.

The plan of the paper is the following. In Section 2 we express the field
equations in terms of the Levi-Civita connection and show that homogeneous and
isotropic models can be interpreted as two-fluid models. In Section 3,
following the method of Coley and Wainwright in \cite{cowajm}, we write the
equations as a two-dimensional dynamical system and analyze the asymptotic
behavior of the models. In Section 4 we consider the two-fluid model resulting
from a modification of the Einstein-Hilbert Lagrangian, cf. (\ref{action}),
and study the possibility of providing a mechanism of accelerating expansion.
Readers unfamiliar with Weyl geometry may find in the Appendix an exposition
of the techniques involved. In this Appendix we also derive the Bianchi
identities and write the Einstein tensor and the Bianchi identities in terms
of the Levi-Civita connection.

\section{Field equations}

We recall that a Weyl space is a manifold endowed with a metric $\mathbf{g}$
and a linear symmetric connection $\mathbf{\nabla}$ which are interrelated
via
\begin{equation}
\nabla_{\mu}g_{\alpha\beta}=-Q_{\mu}g_{\alpha\beta},\label{weyl}%
\end{equation}
where the 1-form $Q_{\mu}$ is customarily called Weyl covariant vector field
(see the Appendix for a full explanation of the notation involved below). We
denote by $D$ the Levi-Civita connection of the metric $g_{\alpha\beta}.$

In \cite{cmq} it was shown that application of the constrained variational
principle to general Lagrangians of the form $L=f\left(  R\right)  ,$ in the
context of Weyl geometry, yields the field equations obtained via the metric
variation in Riemannian spaces with a source tensor depending on the Weyl
vector field. The simplest theory that can be constructed with the constrained
variational principle is obtained from the Lagrangian $L=R.$ The field
equations are (see \cite{cmq})%
\begin{equation}
G_{\left(  \mu\nu\right)  }=-\nabla_{\left(  \mu\right.  }Q_{\left.
\nu\right)  }+Q_{\mu}Q_{\nu}+g_{\mu\nu}\left(  \nabla^{\alpha}Q_{\alpha
}-Q^{\alpha}Q_{\alpha}\right)  =:M_{\mu\nu}.\label{field}%
\end{equation}
If we express the tensor $M_{\mu\nu}$ in terms of the quantities formed with
the Levi-Civita connection $D$ and take into account of (\ref{eins}), the
field equations become
\begin{equation}
\overset{\circ}{G}_{\mu\nu}=\frac{3}{2}\left(  Q_{\mu}Q_{\nu}-\frac{1}{2}%
Q^{2}g_{\mu\nu}\right)  .\label{field1}%
\end{equation}
In the case of integrable Weyl geometry, i.e., when $Q_{\mu}=\partial_{\mu
}\phi,$ the source term is that of a massless scalar field. Taking the
divergence of (\ref{field1}) and using the Bianchi identities (\ref{bianf}) we
conclude that
\begin{equation}
D^{\mu}Q_{\mu}=0.\label{div}%
\end{equation}

In this paper we will be concerned with spatially homogeneous and isotropic
spacetimes. Therefore we have to make the assumption that $Q^{\mu}$ is
hypersurface orthogonal. That means that $Q^{\mu}$ is proportional to the unit
timelike vector field $n^{\mu}$ which is orthogonal to the homogeneous
hypersurfaces,
\[
Q^{\mu}=:qn^{\mu},\;\;\;Q^{2}=Q_{\mu}Q^{\mu}=-q^{2}.
\]
Formally the field equations (\ref{field1}) can be rewritten as
\begin{equation}
\overset{\circ}{G}_{\mu\nu}=\left(  \rho_{1}+p_{1}\right)  n_{\mu}n_{\nu
}+p_{1}g_{\mu\nu},\label{field2}%
\end{equation}
with
\begin{equation}
\rho_{1}=p_{1}=\frac{3}{4}q^{2},\label{ropi}%
\end{equation}
and we see that the equation of state of the $q-$fluid corresponds to stiff
matter. For spatially homogeneous and isotropic models, the field equations
(\ref{field2}) become the system of the equations: the Friedmann
equation\footnote{We adopt the metric and curvature conventions of
\cite{wael}. Here, $a\left(  t\right)  $ is the scale factor, an overdot
denotes differentiation with respect to time $t,$ and units have been chosen
so that $c=1=8\pi G.$}
\[
\left(  \frac{\dot{a}}{a}\right)  ^{2}+\frac{k}{a^{2}}=\frac{1}{4}q^{2},
\]
and the Raychaudhuri equation
\[
2\frac{\ddot{a}}{a}+\left(  \frac{\dot{a}}{a}\right)  ^{2}+\frac{k}{a^{2}%
}=-\frac{3}{4}q^{2}.
\]
Equation (\ref{div}) becomes
\[
\dot{q}-3\frac{\dot{a}}{a}q=0,
\]
which implies that
\[
q=\frac{C}{a^{3}},
\]
where $C$ is a constant. Therefore the energy density and pressure of the
$q-$fluid evolve as $a^{-6}.$

In the following we assume that ordinary matter is described by a perfect
fluid with energy-momentum tensor,
\begin{equation}
T_{\mu\nu}=\left(  \rho_{2}+p_{2}\right)  u_{\mu}u_{\nu}+p_{2}g_{\mu\nu
},\label{pefl}%
\end{equation}
where $u^{\mu}$ denotes the fluid velocity. To preserve the homogeneity and
isotropy of the spacetime it is necessary that
\[
Q^{\mu}=qu^{\mu},\;\;\;Q^{2}=Q_{\mu}Q^{\mu}=-q^{2}.
\]
Therefore we are dealing with a two-fluid model with total energy density and
pressure given by
\begin{equation}
\rho=\rho_{1}+\rho_{2},\;\;\;p=p_{1}+p_{2},\label{rop2}%
\end{equation}
respectively, where
\begin{equation}
p_{1}=\rho_{1},\;\;p_{2}=\left(  \gamma_{2}-1\right)  \rho_{2},\label{state}%
\end{equation}
i.e., $\gamma_{1}=2$ and $\gamma_{2}<\gamma_{1}.$ The field equations take the
final form
\begin{align}
\left(  \frac{\dot{a}}{a}\right)  ^{2}+\frac{k}{a^{2}} &  =\frac{1}{3}\left(
\rho_{1}+\rho_{2}\right)  \label{frie1}\\
\frac{\ddot{a}}{a} &  =-\frac{1}{6}\left[  4\rho_{1}+\left(  3\gamma
_{2}-2\right)  \rho_{2}\right]  \label{rayc1}\\
\dot{\rho}_{1} &  =-6\rho_{1}\frac{\dot{a}}{a}\label{cons1}\\
\dot{\rho}_{2} &  =-3\gamma_{2}\rho_{2}\frac{\dot{a}}{a}.\label{cons2}%
\end{align}
It is well known (see \cite{wael}) that the field equations for one fluid can
be written as a two-dimensional dynamical system for the Hubble variable
$H:=\dot{a}/a$ and the density parameter $\Omega:=\rho/3H^{2}.$ A drawback of
this analysis is that it does not give a complete description of the evolution
for closed models. In fact, at the time of maximum expansion the Hubble
parameter becomes zero and therefore, the time coordinate defined in
\cite{wael} p. 58, cannot be used past the instant of maximum expansion.
Instead, we deal with closed models by defining the compactified density
parameter\emph{ }$\omega,$ (see \cite{wainjm})
\begin{equation}
\Omega=\frac{1}{\tan^{2}\omega},\label{compadjm}%
\end{equation}
or
\[
\omega=\arctan\left(  \frac{\sqrt{3}H}{\sqrt{\rho}}\right)  ,\text{
}\,\;\;\;\text{with \ \ }-\pi/2\leq\omega\leq\pi/2.
\]
We see that $\omega$ is bounded at the instant of maximum expansion ($H=0$)
and also as $\rho\rightarrow0,$ in ever-expanding models.

\section{Phase plane analysis}

We now adopt the Coley and Wainwright formalism for a general model with two
fluids with variable equations of state (cf. \cite{cowajm}). Assuming that
$\gamma_{1}>\gamma_{2}$ and $\gamma_{1}>2/3,$ we define the transition
variable $\chi\in\lbrack-1,1]$
\begin{equation}
\chi=\frac{\rho_{2}-\rho_{1}}{\rho_{2}+\rho_{1}}\label{tran}%
\end{equation}
which describes which fluid is dominant dynamically. The total density
parameter $\Omega=\Omega_{1}+\Omega_{2}$ can be compactified as in
(\ref{compadjm}) and the evolution equation of the variable $\chi$ is obtained
by applying the conservation equation to $\rho_{1}$ and $\rho_{2}.$ Defining a
new time variable $\tau$ by
\[
\frac{d}{dt}=\frac{3\left(  \gamma_{1}-\gamma_{2}\right)  }{2}\sqrt{\frac
{\rho}{3}}\frac{1}{\cos\omega}\frac{d}{d\tau},
\]
and with the same kind of manipulations as in \cite{cowajm}, one obtains the
following dynamical system
\begin{align}
\frac{d\omega}{d\tau} &  =-\frac{1}{2}\left(  b-\chi\right)  \cos2\omega
\cos\omega\nonumber\\
\frac{d\chi}{d\tau} &  =(1-\chi^{2})\sin\omega,\label{2fluidjm}%
\end{align}
where the constant $b$ is
\[
b=\frac{3\left(  \gamma_{1}+\gamma_{2}\right)  -4}{3\left(  \gamma_{1}%
-\gamma_{2}\right)  }>-1.
\]
In our case, we always have $\gamma_{1}=2.$

The phase space of the two-dimensional system (\ref{2fluidjm}) is the closed
rectangle
\[
D=\left[  -\pi/2,\pi/2\right]  \times\left[  -1,1\right]
\]
in the $\omega-\chi$ plane (see Figure 1).%

\begin{figure}[h]
\begin{center}
\includegraphics[
trim=1.418494in 6.293201in 1.186076in 1.765138in,
height=7.005cm,
width=10.8689cm
]%
{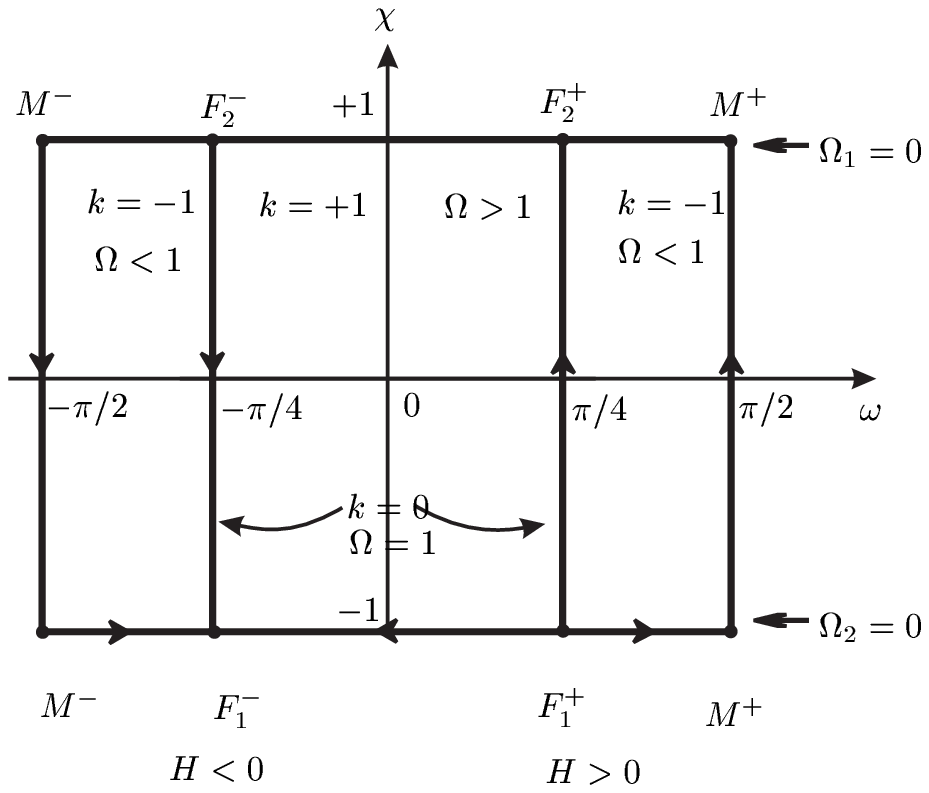}\newline

\end{center}
\caption{The invariant sets and equilibrium points of (\ref{2fluidjm})}.
\label{fig1}%
\end{figure}

By inspection we can see that the line segment $\left\{  \left(  \omega
,\chi\right)  \in D:\omega=\pi/4\right\}  $ is an invariant set of
(\ref{2fluidjm}). It consists of three trajectories, the line segment
\[
\left\{  \left(  \omega,\chi\right)  \in D:\omega=\pi/4,\;-1<\chi<1\right\}
\]
and the equilibrium points $\left(  \pi/4,-1\right)  $ and $\left(
\pi/4,+1\right)  .$ Similarly we can specify the following invariant sets.%

\[%
\begin{array}
[c]{lll}%
\omega=-\pi/2 & \text{contracting empty models} & \Omega=0,H<0\\
\chi=-1 & \text{one-fluid models} & \Omega_{2}=0\\
\chi=+1 & \text{one-fluid models} & \Omega_{1}=0\\
\omega=\pi/4 & \text{expanding flat models} & \Omega=1,H>0\\
\omega=-\pi/4 & \text{contracting flat models} & \Omega=1,H<0\\
\omega=\pi/2 & \text{expanding empty models} & \Omega=0,H>0
\end{array}
\]
It is easy to verify that the equilibrium points lie at the intersection of
these sets. The equilibrium points customarily denoted as $F_{1,2}^{+},$
$F_{1,2}^{-},$ $M^{+},$ $M^{-}$ are
\[%
\begin{array}
[c]{llll}%
F_{1,2}^{+}: & \omega=\pi/4 & \text{expanding flat model} & \Omega
=1,\;k=0,\;H>0\\
F_{1,2}^{-}: & \omega=-\pi/4 & \text{contracting flat model} & \Omega
=1,\;k=0,\;H<0\\
M^{+}: & \omega=\pi/2 & \text{expanding Milne model} & \Omega=0,\;k=-1,\;H>0\\
M^{-}: & \omega=-\pi/2 & \text{contracting Milne model} & \Omega
=0,\;k=-1,\;H<0
\end{array}
\]
and the subscripts indicate which fluid survives. Each of these eight
equilibrium points corresponds to an exact solution of the Einstein equations.

We shall carry out in some detail the stability analysis of the equilibrium
points of (\ref{2fluidjm}) in the case $\gamma_{2}=1,$ corresponding to dust.
It turns out that linearization of (\ref{2fluidjm}) is sufficient to determine
the global phase portrait of the system. In fact, the derivative matrix
$J\left(  \omega,\chi\right)  $ of the vector field of (\ref{2fluidjm}) is
non-singular and, $J$ computed at each of the eight equilibrium points, has
two real eigenvalues. Therefore, the Hartman-Grobman theorem applies in the
case of (\ref{2fluidjm}). It is easy to verify that $J$ computed at all
equilibrium points is diagonal. Therefore, we conclude in a straightforward
manner that $\left(  -\pi/2,-1\right)  ,\left(  -\pi/4,+1\right)  ,\left(
\pi/4,+1\right)  ,\left(  \pi/2,-1\right)  $ are saddle points, $\left(
-\pi/2,+1\right)  ,\left(  \pi/4,-1\right)  $ are unstable nodes and $\left(
-\pi/4,-1\right)  ,\left(  \pi/2,+1\right)  ,$ are stable nodes. The phase
portrait is shown in Figure 2.%

\begin{figure}[h]
\begin{center}
\includegraphics[
trim=1.974312in 6.030010in 1.528499in 2.133606in,
height=6.8073cm,
width=9.1555cm
]%
{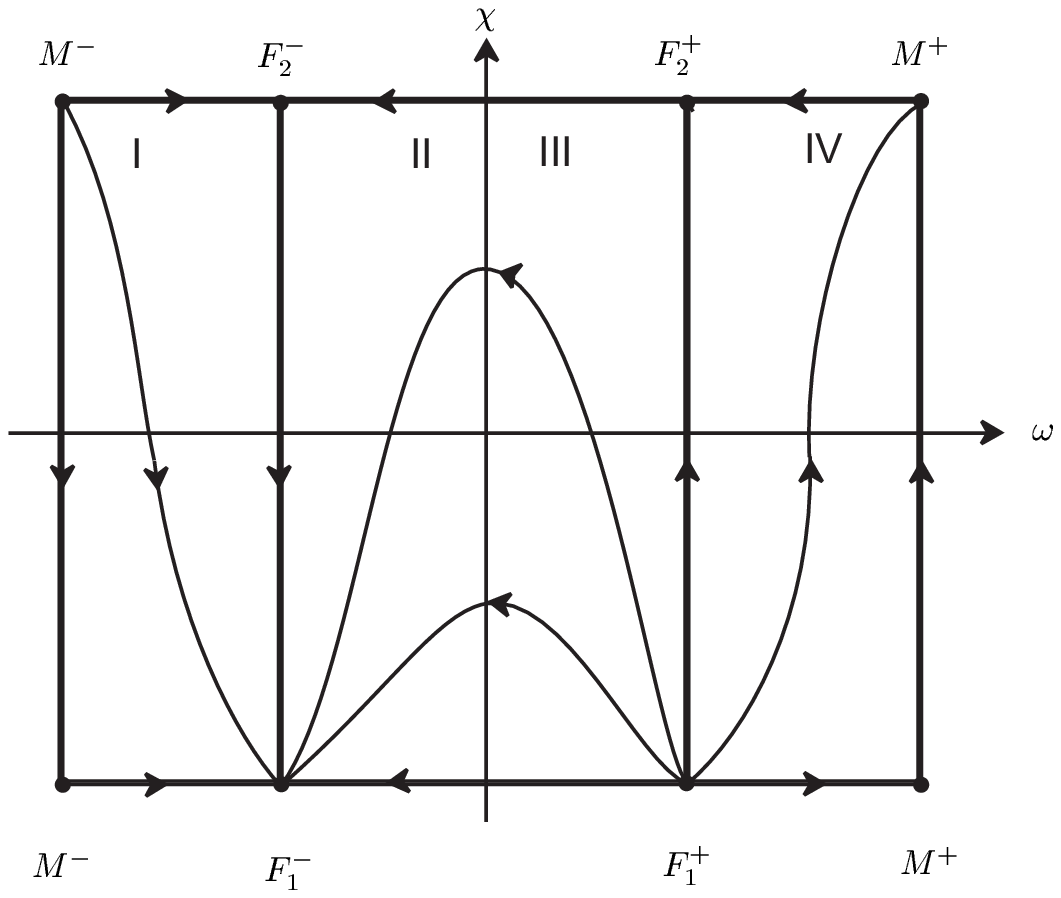}\newline
\end{center}
\caption{The phase portrait of (\ref{2fluidjm}) with $\gamma_{2}=1.$}
\label{fig2}%
\end{figure}

Regions III and IV correspond to expanding models. The $F_{1}^{+}$ is a past
attractor of all models with $\Omega>0$, i.e., the evolution near the big bang
is approximated by the flat FRW model where the Weyl fluid dominates. Open
models expand indefinitely and the evolution is approximated by the Milne
universe at late time. Flat models expand indefinitely and the evolution is
approximated by the flat FRW universe at late time. In both cases the ``real''
second fluid dominates at late times while the $q-$fluid becomes
insignificant. On the other hand, any initially expanding closed model in
region III, however close to $F_{2}^{+}$, eventually recollapses and the
evolution is approximated by the flat FRW model where the Weyl fluid dominates.

In the case $\gamma_{2}=0$ corresponding to a positive cosmological constant,
a new equilibrium point $\left(  0,2/3\right)  $ appears, denoted by $E$ (see
Figure 3). It corresponds to the Einstein static model.
\begin{figure}[h]
\begin{center}
\includegraphics[
trim=1.580607in 6.031180in 1.597149in 2.118399in,
height=6.8293cm,
width=9.775cm
]%
{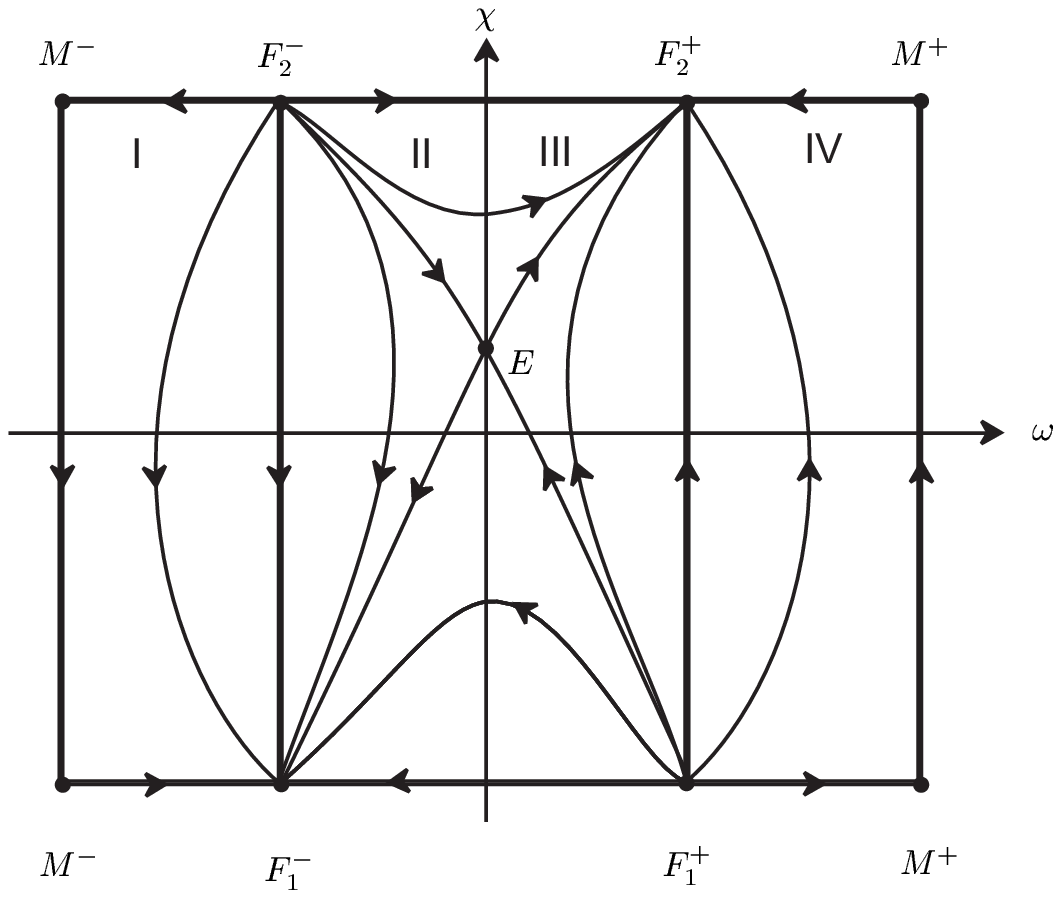}\newline
\end{center}
\caption{The phase portrait of (\ref{2fluidjm}) with $\gamma_{2}=0.$}
\label{fig3}%
\end{figure}
Consider for example a closed model in region III starting close to the
$F_{1}^{+}$ model. Its trajectory passes close to the Einstein model,
indicating a phase of halted expansion and asymptotically approaches the de
Sitter model $F_{2}^{+}$. Similarly, an open model in region IV starting close
to $F_{1}^{+}$ asymptotically approaches the de Sitter model. This attracting
property of the de Sitter solution for all expanding models is not restricted
only to isotropic cosmology. In fact, the cosmic no-hair conjecture states
that all expanding universe models with a positive cosmological constant,
asymptotically approach the de Sitter solution (see for example \cite{wald}).

We therefore conclude that the Weyl fluid has significant contribution only
near the cosmological singularities. In expanding models the ``real'' fluid
always dominates at late times and therefore the contribution of the Weyl
fluid to the total energy-momentum tensor is important only at early times.

\section{Extensions}

The field equations (\ref{field1}) constitute the generalization of the
Einstein equations in a Weyl spacetime in the sense that they come from the
Lagrangian $L=R.$ There is however an alternative view, namely that the pair
$\left(  Q,\mathbf{g}\right)  $ which defines the Weyl spacetime also enters
into the gravitational theory and therefore, the field $Q$ must be contained
in the Lagrangian independently from $\mathbf{g.}$ In the case of integrable
Weyl geometry, i.e. when $Q_{\mu}=\partial_{\mu}\phi$ where $\phi$ is a scalar
field, the pair $\left(  \phi,g_{\mu\nu}\right)  $ constitute the set of
fundamental geometrical variables. A simple Lagrangian involving this set is
given by
\begin{equation}
L=R+\xi\nabla^{\mu}Q_{\mu}, \label{action}%
\end{equation}
where $\xi$ is a constant. Motivations for considering theory (\ref{action})
see can be found in \cite{nove,sasa} (see also \cite{kome} for a
multidimensional approach and \cite{oss} for an extension of (\ref{action}) to
include an exponential potential function of $\phi$). By varying the action
corresponding to (\ref{action}) with respect to both $g_{\mu\nu}$ and $\phi$
one obtains
\begin{equation}
\overset{\circ}{G}_{\mu\nu}=\frac{4\xi-3}{2}\left(  Q_{\mu}Q_{\nu}-\frac{1}%
{2}Q^{2}g_{\mu\nu}\right)  ,\;\;\;\;\;\text{and\ \ \ \ }\overset{\circ
}{\square}\phi=0.\text{\ } \label{nove}%
\end{equation}
Note that the second of (\ref{nove}) comes from the variational procedure
while (\ref{div}) is a consequence of the Bianchi identities.

For isotropic cosmologies one can again interpret the source term in
(\ref{nove}) as a perfect fluid with density and pressure given by
\begin{equation}
\rho_{1}=p_{1}=\lambda q^{2},\,\;\;\;\;\;\;\lambda=\frac{4\xi-3}{2}
\label{lamd}%
\end{equation}
respectively. However, if we allow for $\xi$ to be a free parameter, equation
(\ref{lamd}) implies that the energy density and pressure of the $q-$fluid may
take negative values. An important consequence is that open models ($k=-1$) in
vacuum with $\lambda<0$ avoid the initial singularity \cite{nove}.

In the following we assume that $\lambda<0$ and apply again the two-fluid
analysis of Section 3. A sufficient negative contribution of the $q-$fluid to
the total energy density and pressure, cf. (\ref{rop2}), may provoke an
accelerating expansion of the universe. In fact, the $\rho+3p$ term in the
Raychaudhuri equation%
\[
\frac{\ddot{a}}{a}=-\frac{1}{6}\left(  \rho+3p\right)  ,
\]
becomes negative provided that $4\rho_{1}<\left(  2-3\gamma_{2}\right)
\rho_{2}.$ Unfortunately, this cannot explain the observed acceleration of the
Universe, as we shall see in a moment.

The Friedmann constraint (\ref{frie1}) implies that the total density
parameter $\Omega=\Omega_{1}+\Omega_{2}$ is still non-negative for flat and
closed models, but $\Omega$ may be negative for open models. We conclude that
$\Omega$ cannot be compactified as in (\ref{compadjm}) and the transition
variable $\chi$ defined by (\ref{tran}) is unbounded. Therefore we cannot
apply the Coley and Wainwright formalism developed in Section 4.

Nevertheless, we can infer about the asymptotic behavior of the system by
looking at the original field equations, (\ref{frie1})-(\ref{cons2}). From
these equations we see that the state $\left(  a,\dot{a},\rho_{1},\rho
_{2}\right)  \in\mathbb{R}^{4}$ of the system lies on the hypersurface defined
by the constraint (\ref{frie1}) and the remaining evolution equations can be
written as a constrained four-dimensional dynamical system. By standard
arguments one can show that for flat and open models the sign of $H$ is
invariant (see for example \cite{miri}), therefore an initially expanding
universe remains ever expanding. Suppose that $\ddot{a}\left(  t_{0}\right)
>0$ at some time $t_{0}.$ Then no solution of (\ref{rayc1})-(\ref{cons2})
exists such that $\ddot{a}\left(  t\right)  >0$ for all $t>t_{0}.$ In fact,
(\ref{cons1}) and (\ref{cons2}) can be solved to give%
\begin{equation}
\rho_{1}=\frac{C_{1}}{a^{6}},\;\;\;\;\rho_{2}=\frac{C_{2}}{a^{3\gamma_{2}}%
},\label{rhos}%
\end{equation}
where $C_{1}$ and $C_{2}$ are constants. Since $\gamma_{2}<2,$ the ``real''
second fluid in an expanding universe eventually dominates. Therefore, the
term $4\rho_{1}+\left(  3\gamma_{2}-2\right)  \rho_{2}$ in (\ref{rayc1})
becomes positive at some time $t_{1}>t_{0}$ and evidently $\ddot{a}\left(
t\right)  <0$ for all $t>t_{1}.$ We conclude that even if the universe expands
initially with acceleration, eventually evolves according to the Friedmann cosmology.

Note that for flat and open models there exist solutions without an initial
singularity. To see this, consider for example $\gamma_{2}=1$ and substitute
(\ref{rhos}) into the Friedmann equation (\ref{frie1}) with $k=-1$. Then
(\ref{frie1}) can be interpreted as describing the motion of a particle with
total energy $E=1/2$ in the effective potential%
\begin{equation}
V\left(  a\right)  =-\frac{A}{a}+\frac{B}{a^{4}},\;\;\;\;A,B>0.\label{pote}%
\end{equation}%
\begin{figure}[h]
\begin{center}
\includegraphics[
trim=1.689786in 6.821923in 2.004915in 1.952296in,
height=5.6409cm,
width=8.7887cm
]%
{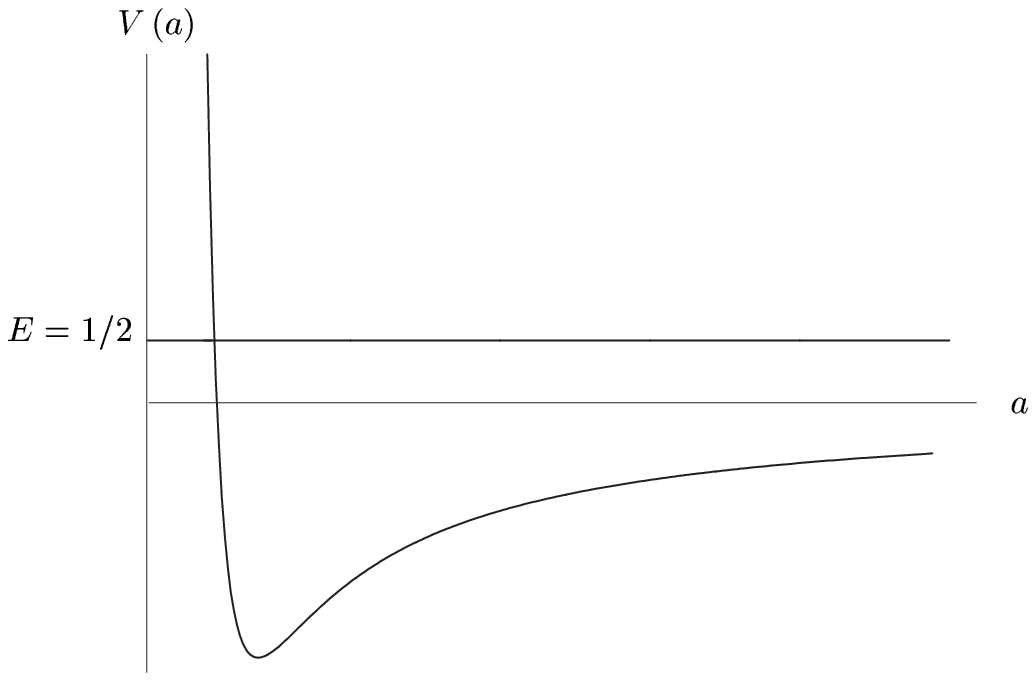}\newline
\end{center}
\caption{The effective potential (\ref{pote}).}
 \label{fig4}
\end{figure}
In Figure 4,
we see that motion is impossible for values of the scale factor smaller than some
minimum $a_{\min}$ and therefore, $a\left(  t\right)  \geq a_{\min}$ for all $t.$ A
similar argument shows that flat models, also avoid the initial singularity, a
conclusion already obtained using phase portrait analysis by Oliveira \emph{et al}
\cite{oss}.

To summarize, the extension to Weyl geometry in theory (\ref{action}) cannot
explain the observed acceleration of the Universe. Since the real fluid
dominates at late times, the accelerated expansion due to the Weyl fluid is
important only at early times. Nevertheless, it is possible that theory
(\ref{action}) could provide a geometric explanation of an inflationary phase
present in the early universe.

\section*{Acknowledgements}

I am grateful to S Cotsakis for reading the manuscript and commenting on this
work. \renewcommand{\theequation}{A.\arabic
{equation}}

\section*{Appendix: Weyl geometry}

The connection $\mathbf{\nabla}$ of the Weyl space is determined by the
symmetric functions
\[
\Gamma_{\beta\gamma}^{\alpha}=\left\{  _{\beta\gamma}^{\alpha}\right\}
+C_{\beta\gamma}^{\alpha},
\]
where $\left\{  _{\beta\gamma}^{\alpha}\right\}  $ are the Levi-Civita
connection coefficients and the tensor field $C_{\beta\gamma}^{\alpha}$ is
given by
\[
C_{\beta\gamma}^{\alpha}=\frac{1}{2}\left(  \delta_{\beta}^{\alpha}Q_{\gamma
}+\delta_{\gamma}^{\alpha}Q_{\beta}-g_{\beta\gamma}Q^{\alpha}\right)  .
\]
Given a metric and a connection satisfying (\ref{weyl}), \emph{viz.}
$\nabla_{\mu}g_{\alpha\beta}=-Q_{\mu}g_{\alpha\beta},$ then for every
differentiable function $\sigma,$ the metric and the 1-form defined by%
\begin{equation}
g_{\alpha\beta}^{\prime}=\sigma g_{\alpha\beta},\;\;\;\;\;Q_{\mu}^{\prime
}=Q_{\mu}-\partial_{\mu}\ln\sigma\label{gauge}%
\end{equation}
respectively, also satisfy (\ref{weyl}). Thus, $g_{\alpha\beta}$ and $Q_{\mu}$
are far from unique; rather $\mathbf{g}$ belongs to an equivalence class
$\left[  g\right]  $ of metrics and for each $\mathbf{g}\in\left[  g\right]
,$ there exists a unique 1-form $Q$ such that (\ref{weyl}) is satisfied. A
particular choice of a pair $\left(  Q,\mathbf{g}\right)  $ is called a gauge
and (\ref{gauge}) is a gauge transformation.

The Riemann tensor is defined by
\begin{equation}
R^{\alpha}{}_{\beta\gamma\delta}=\partial_{\gamma}\Gamma_{\beta\delta}%
^{\alpha}-\partial_{\delta}\Gamma_{\beta\gamma}^{\alpha}+\Gamma_{\sigma\gamma
}^{\alpha}\Gamma_{\beta\delta}^{\sigma}-\Gamma_{\sigma\delta}^{\alpha}%
\Gamma_{\beta\gamma}^{\sigma}. \label{riem}%
\end{equation}
In contrast to the familiar property in Riemannian geometry, in Weyl geometry
$R_{\mu\nu\alpha\beta}$ is not antisymmetric in its first two indices; it
satisfies%
\begin{equation}
2R_{\left(  \mu\nu\right)  \alpha\beta}=g_{\mu\nu}H_{\alpha\beta},
\label{riem1}%
\end{equation}
where
\begin{equation}
H_{\alpha\beta}:=\partial_{\alpha}Q_{\beta}-\partial_{\beta}Q_{\alpha}.
\label{curl}%
\end{equation}
Note that $H_{\alpha\beta}$ is the same for all derivative operators, for
example $H_{\alpha\beta}=\nabla_{\alpha}Q_{\beta}-\nabla_{\beta}Q_{\alpha}.$
When first introduced by Weyl, $H_{\alpha\beta}$ was supposed to represent the
Faraday 2-form. The relation $2R_{\left(  \mu\nu\right)  \alpha\beta}%
=g_{\mu\nu}H_{\alpha\beta}$ can be obtained either directly from (\ref{riem}),
or more quickly by applying on the metric the anticommutativity property of
$\nabla_{\alpha}$:%
\[
\left(  \nabla_{\alpha}\nabla_{\beta}-\nabla_{\beta}\nabla_{\alpha}\right)
g_{\mu\nu}=R^{\rho}{}_{\mu\beta\alpha}g_{\rho\nu}+R^{\rho}{}_{\nu\beta\alpha
}g_{\mu\rho}.
\]
The Ricci tensor is defined by
\[
R_{\alpha\beta}=R^{\mu}{}_{\alpha\mu\beta}%
\]
and is not a symmetric tensor as we are accustomed to in Riemannian geometry;
its antisymmetric part is given by
\[
R_{\left[  \alpha\beta\right]  }=H_{\alpha\beta}.
\]
The geometric meaning of (\ref{riem1}) is that the measuring units change from
point to point (see \cite{eddi} for a discussion). In fact, the length
$l^{2}=g_{\mu\nu}V^{\mu}V^{\nu}$ of a vector field $V$ is not invariant when
$V$ is parallely translated in a small circuit ; instead, using the standard
argument one may show that for a closed circuit enclosing an elementary area
$\delta S^{\alpha\beta},$ the variation of the length is
\[
\delta\left(  l^{2}\right)  =R_{\left(  \mu\nu\right)  \alpha\beta}V^{\mu
}V^{\nu}\delta S^{\alpha\beta}=l^{2}H_{\alpha\beta}\delta S^{\alpha\beta}.
\]
As a consequence, there is an additional loss of synchronization for two
initially synchronized clocks following different paths from one point to
another, due to the distinct variation of the units of measure along the two
paths. This is the so-called second clock effect \cite{perl}.

When the Weyl vector field is the gradient of a scalar function, then the curl
$H_{\alpha\beta}$ vanishes identically and we have the so-called integrable
Weyl geometry. In that case the spacetime is not a genuine Weyl space, but a
conformally equivalent Riemann space. In fact, if $Q_{\mu}=\partial_{\mu}%
\phi,$ where $\phi$ is a scalar field, it is easy to see that it can be gauged
away by the conformal transformation $\widetilde{g}_{\alpha\beta}=\left(
\exp\phi\right)  g_{\alpha\beta}$ and therefore the original space is not a
general Weyl space, but a Riemann space with an undetermined gauge
\cite{scho}. However, most studies of gravity theories were developed in the
framework of integrable Weyl geometry in which the second clock effect is
eliminated \cite{nove,kome,sasa,oss}.

\subsection*{ Bianchi identities}

For the convenience of the reader we present a few identities. We denote by
$D$ the Levi-Civita connection.\medskip

1. $\nabla^{\alpha}Q_{\alpha}+Q^{2}=\nabla_{\alpha}Q^{\alpha}$

2. $\square:=g^{\alpha\beta}\nabla_{\alpha}\nabla_{\beta}=\nabla^{\alpha
}\nabla_{\alpha}=\nabla_{\alpha}\nabla^{\alpha}-Q_{\alpha}\nabla^{\alpha}$

3. $Q_{\alpha}\nabla^{\alpha}=Q^{\alpha}\nabla_{\alpha}$

4. $\nabla_{\mu}A^{\nu}=D_{\mu}A^{\nu}+C_{\mu\alpha}^{\nu}A^{\alpha
},\;\;\nabla_{\mu}B_{\nu}=D_{\mu}B_{\nu}-C_{\mu\nu}^{\alpha}B_{\alpha}$

5. $C_{\alpha\mu}^{\alpha}=2Q_{\mu},\;\;\;g^{\mu\nu}C_{\mu\nu}^{\alpha
}=-Q^{\alpha}$

6. $\nabla_{\mu}Q_{\nu}=D_{\mu}Q_{\nu}-Q_{\mu}Q_{\nu}+\frac{1}{2}Q^{2}%
g_{\mu\nu}$

7. $\nabla^{\alpha}Q_{\alpha}=D^{\alpha}Q_{\alpha}+Q^{2},\;\;\;\nabla_{\alpha
}Q^{\alpha}=D_{\alpha}Q^{\alpha}+2Q^{2}$

8. $\nabla_{\alpha}Q^{2}=Q_{\alpha}Q^{2}+2Q^{\mu}\nabla_{\alpha}Q_{\mu}%
=Q^{\mu}\nabla_{\alpha}Q_{\mu}+Q_{\mu}\nabla_{\alpha}Q^{\mu}$

9. $\left(  \nabla^{\mu}\nabla_{\alpha}-\nabla_{\alpha}\nabla^{\mu}\right)
Q_{\mu}=Q_{\mu}R^{\mu}{}_{\alpha}+Q^{\mu}H_{\alpha\mu}=Q_{\mu}R\underline
{^{\mu}{}_{\alpha}}\medskip$

We now define the auxiliary operator $\overline{\nabla}_{\alpha}%
:=\nabla_{\alpha}-Q_{\alpha};$ thus $\overline{\nabla}_{\alpha}\ $commutes
with $g^{\mu\nu}.$ Note that $\overline{\nabla}_{\alpha}$ is not a derivative
operator. For every torsion-free connection, the Bianchi identities take the
form
\[
\nabla_{\rho}R^{\alpha}{}_{\mu\sigma\nu}+\nabla_{\nu}R^{\alpha}{}_{\mu
\rho\sigma}+\nabla_{\sigma}R^{\alpha}{}_{\mu\nu\rho}=0.
\]
We contract on $\alpha-\sigma$ to obtain%
\[
\nabla_{\rho}R_{\mu\nu}-\nabla_{\nu}R{}_{\mu\rho}+\nabla_{\alpha}R^{\alpha}%
{}_{\mu\nu\rho}=0.
\]
Transvection with $g^{\rho\mu}$ yields%
\[
\overline{\nabla}_{\mu}R^{\mu}{}_{\nu}-\overline{\nabla}_{\nu}R+\overline
{\nabla}_{\alpha}R^{\alpha\mu}{}_{\nu\mu}=0\Rightarrow\overline{\nabla}_{\mu
}R^{\mu}{}_{\nu}-\overline{\nabla}_{\nu}R-\overline{\nabla}_{\alpha}\left(
g^{\alpha\mu}{}H_{\mu\nu}\right)  +\overline{\nabla}_{\alpha}R^{\mu\alpha}%
{}_{\mu\nu}=0
\]%
\begin{equation}
\Rightarrow2\overline{\nabla}_{\mu}R^{\mu}{}_{\nu}-\overline{\nabla}_{\nu
}R-\overline{\nabla}_{\alpha}\left(  g^{\alpha\mu}{}H_{\mu\nu}\right)
=0.\label{2}%
\end{equation}
On the other hand,
\[
R^{\mu}{}_{\nu}=g^{\mu\alpha}R_{\alpha\nu}=g^{\mu\alpha}\left(  R_{\left(
\alpha\nu\right)  }+H_{\alpha\nu}\right)  =:R\underline{^{\mu}{}_{\nu}}%
+g^{\mu\alpha}H_{\alpha\nu},
\]
hence equation (\ref{2}) yields%
\[
2\overline{\nabla}_{\mu}R\underline{^{\mu}{}_{\nu}}-\overline{\nabla}_{\nu
}R+\overline{\nabla}_{\alpha}\left(  g^{\alpha\mu}{}H_{\mu\nu}\right)  =0
\]
and since $\overline{\nabla}_{\alpha}\ $commutes with $g^{\alpha\mu}$ we
obtain%
\begin{equation}
\overline{\nabla}_{\mu}\left(  R\underline{^{\mu}{}_{\nu}}-\frac{1}{2}%
\delta_{\nu}^{\mu}R\right)  =-\frac{1}{2}\overline{\nabla}^{\mu}H_{\mu\nu
}.\label{bian}%
\end{equation}
Denoting by $G\underline{^{\mu}{}_{\nu}}$ the mixed tensor corresponding to
the symmetric part of the Einstein tensor, i.e.,
\[
G\underline{^{\mu}{}_{\nu}}=g^{\mu\alpha}G_{\left(  \alpha\nu\right)  },
\]
the Bianchi identities (\ref{bian}) can be finally written as
\begin{equation}
\left(  \nabla_{\mu}-Q_{\mu}\right)  G\underline{^{\mu}{}_{\nu}}=-\frac{1}%
{2}\left(  \nabla^{\mu}-Q^{\mu}\right)  H_{\mu\nu}.\label{bianchi}%
\end{equation}
Note that in the case of integrable Weyl geometry, equation (\ref{bianchi})
reduces to $\nabla_{\mu}G\underline{^{\mu}{}_{\nu}}=Q_{\mu}G\underline{^{\mu
}{}_{\nu}}$ which is usually referred in the literature as Bianchi identity
(cf. \cite{haba}).

\subsection*{Relation with the Levi-Civita connection}

It is useful to express the Einstein tensor and the Bianchi identities in
terms of quantities formed with the Levi-Civita connection, $D$. Starting with
the definition of the Riemann tensor we arrive at%
\begin{align*}
R^{\alpha}{}_{\beta\gamma\delta}  &  =\overset{\circ}{R^{\alpha}}_{\beta
\gamma\delta}+\partial_{\gamma}C_{\beta\delta}^{\alpha}-\partial_{\delta
}C_{\beta\gamma}^{\alpha}+C_{\mu\gamma}^{\alpha}C_{\beta\delta}^{\mu}%
-C_{\mu\delta}^{\alpha}C_{\beta\gamma}^{\mu}\\
&  +\left\{  _{\mu\gamma}^{\alpha}\right\}  C_{\beta\delta}^{\mu}+\left\{
_{\beta\delta}^{\mu}\right\}  C_{\mu\gamma}^{\alpha}-\left\{  _{\mu\delta
}^{\alpha}\right\}  C_{\beta\gamma}^{\mu}-\left\{  _{\beta\gamma}^{\mu
}\right\}  C_{\mu\delta}^{\alpha},
\end{align*}
where the accent $\overset{\circ}{}$ denotes a quantity formed with the
Levi-Civita connection. We contract on $\alpha,\gamma$ and take account of
$C_{\alpha\mu}^{\alpha}=2Q_{\mu},\;\;g^{\mu\nu}C_{\mu\nu}^{\alpha}=-Q^{\alpha
},\;\;\left\{  _{\alpha\mu}^{\alpha}\right\}  =\partial_{\mu}\ln\sqrt{-g}$ to
obtain for the symmetric part of the Ricci tensor,%
\[
R_{\left(  \beta\delta\right)  }=\overset{\circ}{R}_{\beta\delta}-D_{\left(
\beta\right.  }Q_{\left.  \delta\right)  }+\frac{1}{2}Q_{\beta}Q_{\delta
}-\frac{1}{2}g_{\beta\delta}\left(  D^{\mu}Q_{\mu}+Q^{2}\right)
\]
and upon a new contraction,%
\[
R=\overset{\circ}{R}-3D_{\alpha}Q^{\alpha}-\frac{3}{2}Q^{2}.
\]
Therefore the Einstein tensor takes the form
\begin{equation}
G_{\left(  \alpha\beta\right)  }=\overset{\circ}{G}_{\alpha\beta}-D_{\left(
\alpha\right.  }Q_{\left.  \beta\right)  }+\frac{1}{2}Q_{\alpha}Q_{\beta
}+g_{\alpha\beta}\left(  D^{\mu}Q_{\mu}+\frac{1}{4}Q^{2}\right)  .
\label{eins}%
\end{equation}

We now turn to the Bianchi identities. We calculate the right-hand side (RHS)
of (\ref{bianchi})
\begin{equation}
\left(  \nabla_{\alpha}-Q_{\alpha}\right)  G\underline{^{\alpha}{}_{\nu}%
}=D_{\alpha}G\underline{^{\alpha}{}_{\nu}}+C_{\alpha\beta}^{\alpha}%
G\underline{^{\beta}{}_{\nu}}-C_{\alpha\nu}^{\beta}G\underline{^{\alpha}%
{}_{\beta}}-Q_{\alpha}G\underline{^{\alpha}{}_{\nu}}. \label{bian1}%
\end{equation}
Taking into account of (\ref{eins}) the first term in the RHS of (\ref{bian1})
can be written as
\begin{multline*}
D_{\alpha}G\underline{^{\alpha}{}_{\nu}}=\\
D_{\alpha}\overset{\circ}{G^{\alpha}}_{\nu}-\frac{1}{2}\overset{\circ}%
{\square}Q_{\nu}-\frac{1}{2}D_{\alpha}D_{\nu}Q^{\alpha}+D_{\nu}D_{\alpha
}Q^{\alpha}+\frac{1}{2}Q_{\nu}D_{\alpha}Q^{\alpha}+\frac{1}{2}\left(
Q^{\alpha}D_{\alpha}\right)  Q_{\nu}+\frac{1}{4}D_{\nu}Q^{2}.
\end{multline*}
We use the anticommutativity of the derivative operator $D_{\alpha}$ to write
the terms $-\frac{1}{2}D_{\alpha}D_{\nu}Q^{\alpha}+D_{\nu}D_{\alpha}Q^{\alpha
}$ as
\[
-Q_{\alpha}\overset{\circ}{R^{\alpha}}_{\nu}+\frac{1}{2}D_{\alpha}D_{\nu
}Q^{\alpha}.
\]
The second and the last terms in the RHS of (\ref{bian1}) can be written as%
\begin{multline*}
C_{\alpha\beta}^{\alpha}G\underline{^{\beta}{}_{\nu}}-Q_{\alpha}%
G\underline{^{\alpha}{}_{\nu}}=Q_{\alpha}G\underline{^{\alpha}{}_{\nu}}=\\
Q_{\alpha}\overset{\circ}{G^{\alpha}}_{\nu}-\frac{1}{2}\left(  Q^{\alpha
}D_{\alpha}\right)  Q_{\nu}-\frac{1}{4}D_{\nu}Q^{2}+\frac{1}{2}Q_{\nu}%
Q^{2}+Q_{\nu}D_{\alpha}Q^{\alpha}+\frac{1}{4}Q_{\nu}Q^{2}.
\end{multline*}
The third term in the RHS of (\ref{bian1}) can be written as%
\begin{align*}
-C_{\alpha\nu}^{\beta}G\underline{^{\alpha}{}_{\beta}}  &  =-\frac{1}%
{2}\left(  \delta_{\alpha}^{\beta}Q_{\nu}+\delta_{\nu}^{\beta}Q_{\alpha
}-g_{\alpha\nu}Q^{\beta}\right)  G\underline{^{\alpha}{}_{\beta}}=\frac{1}%
{2}Q_{\nu}R=\\
&  \frac{1}{2}Q_{\nu}\overset{\circ}{R}-\frac{3}{2}Q_{\nu}D_{\alpha}Q^{\alpha
}-\frac{3}{4}Q_{\nu}Q^{2}.
\end{align*}
Putting all these together we obtain from equation (\ref{bian1})%
\[
\left(  \nabla_{\alpha}-Q_{\alpha}\right)  G\underline{^{\alpha}{}_{\nu}%
}=-\frac{1}{2}\overset{\circ}{\square}Q_{\nu}+\frac{1}{2}D_{\alpha}D_{\nu
}Q^{\alpha}.
\]
On the other hand,%
\begin{align*}
\left(  \nabla^{\alpha}-Q^{\alpha}\right)  H_{\alpha\nu}  &  =g^{\alpha\mu
}\left(  D_{\mu}H_{\alpha\nu}-C_{\mu\alpha}^{\rho}H_{\rho\nu}-C_{\mu\nu}%
^{\rho}H_{\alpha\rho}\right)  -Q^{\alpha}H_{\alpha\nu}=\\
&  \overset{\circ}{\square}Q_{\nu}-D_{\alpha}D_{\nu}Q^{\alpha}-Q^{\alpha
}H_{\alpha\nu}.
\end{align*}
Therefore the Bianchi identities (\ref{bianchi}) reduce to
\begin{equation}
Q^{\alpha}H_{\alpha\nu}=0. \label{bianf}%
\end{equation}

\end{document}